\documentclass[a4paper,10pt,twocolumn]{article}
\usepackage{amssymb}
\usepackage{textcomp}
\usepackage{bm}
\usepackage[super]{cite}

\usepackage{graphicx}
\usepackage[font={sf,small},labelfont=bf,labelsep=period,belowskip=-15pt]{caption}
\usepackage[small,compact]{titlesec}
\usepackage{flushend}
\usepackage{pict2e}
\usepackage{textgreek}
\usepackage[T1]{fontenc}
\usepackage{lmodern} 
\usepackage{subcaption}
\usepackage{url}

\usepackage{amsmath}
\usepackage{amsthm}
\usepackage{amsfonts}
\usepackage{color}
\usepackage[table,xcdraw]{xcolor} 
\graphicspath{{../}}
\usepackage{xr}  		
\author{}

\makeatletter
\renewcommand\@citess[1]{\textsuperscript{[#1]}}

\usepackage{soul,xcolor}
\setstcolor{blue}
\newcommand*{\rem}[1]{}
\newcommand*{\ins}[1]{#1}

\begin{document}
\title{\flushleft{\textsf{Surface Reconstruction Limited Conductivity in Block-Copolymer Li Battery Electrolytes 
 \\
\large{\textit{Preston Sutton, Peter Bennington, Shrayesh N. Patel, Morgan Stefik, Ulrich B. Wiesner, Paul F. Nealey, Ullrich Steiner, Ilja Gunkel}}}
}}
\date{\vspace{-1.8cm}}
\twocolumn[
  \begin{@twocolumnfalse}
    \maketitle
    \begin{abstract}
    \noindent
    \textsf{Solid polymer electrolytes for lithium batteries promise improvements in safety and energy density if their conductivity can be increased. Nanostructured block copolymer electrolytes specifically have the potential to provide both good ionic conductivity and good mechanical properties. This study shows that the previously neglected nanoscale composition of the polymer electrolyte close to the electrode surface has an important effect on impedance measurements, despite its negligible extent compared to the bulk electrolyte. Using standard stainless steel blocking electrodes, the impedance of lithium salt-doped poly(isoprene-\emph{b}-styrene-\emph{b}-ethylene oxide) (ISO) exhibited a marked decrease upon thermal processing of the electrolyte. In contrast, covering the electrode surface with a low molecular weight poly(ethylene oxide) (PEO) brush resulted in higher and more reproducible conductivity values, which were insensitive to the thermal history of the device. A qualitative model of this effect is based on the hypothesis that ISO surface reconstruction at the different electrode surfaces leads to a change in the electrostatic double layer, affecting electrochemical impedance spectroscopy measurements. As a main result, PEO-brush modification of electrode surfaces is beneficial for the robust electrolyte performance of PEO-containing block-copolymers and may be crucial for their accurate characterization and use in Li-ion batteries.}

\bigskip

    \end{abstract}
  \end{@twocolumnfalse}
]

\noindent

\let\thefootnote\relax\footnotetext{\vspace{-3mm}}
\footnotetext{\hspace{-6.3mm}
\noindent \textsf{P.\ Sutton, Prof.\ U. Steiner, Dr.\ I. Gunkel\\
Adolphe Merkle Institute, Ch.\ des Verdiers 4, 1700 Fribourg, Switzerland\\
E-mail: ilja.gunkel@unifr.ch\\}

\vspace{-2mm}\noindent
\textsf{P.\ Bennington, Prof.\ S. N. Patel\textsuperscript{[$\dagger$]}, Prof.\ P. F. Nealey\textsuperscript{[$\dagger$]}\\
Pritzker School of Molecular Engineering, The University of Chicago, 5640 S.\ Ellis Avenue, Argonne, IL 60437, USA\\}

\vspace{-2mm}\noindent
\textsf{\textsuperscript{[$\dagger$]}Chemical Sciences and Engineering Division and Materials Science Division, Argonne National Laboratory, Lemont, Illinois 60439, USA\\}

\vspace{-2mm}\noindent
\textsf{Prof.\ M. Stefik\textsuperscript{[$\ddagger$]}, Prof.\ U. B. Wiesner\\
Department of Materials Science and Engineering, Cornell University, 214 Bard Hall, Ithaca, NY 14853-1501, USA\\}

\vspace{-2mm}\noindent
\textsf{\textsuperscript{[$\ddagger$]}Department of Chemistry and Biochemistry, University of South Carolina, Columbia, SC 29208, USA}}

\section{Introduction}
Solid polymer electrolytes (SPEs) are a potentially disruptive advance in lithium battery technology\cite{Manthiram2017,Janek2016}. In contrast to other electrolyte systems, like liquids, gelled polymers, and inorganic solids, the benefits of SPEs include relatively low volatility (increased safety), compatibility with metallic lithium and dendrite resistance (increased energy density), and relatively low-cost materials and manufacturing\cite{Mindemark2018}. 

The major challenge in designing SPEs is their low ionic conductivity for materials that are stiff enough to realize these benefits. Unfortunately, the requisite mechanical properties are accompanied by restricted polymer chain motion which limits ionic conductivity\cite{Ratner2000}. Thus, the crux of the development challenge is finding new strategies to decouple mechanical properties from ion-conductivity.  

Poly(ethylene oxide) (PEO) is a common constituent of polymer electrolytes in solid-state electrochemical devices, where homopolymers can readily meet minimum battery conductivity requirements of $10^{-4}$\,S\,cm$^{-1}$\cite{Long2016}. However, since Li conduction occurs predominantly in the amorphous PEO phase, except for specific cases\cite{Gadjourova2001}, the high conductivity values are normally achieved above the PEO melting temperature, where these polymers form a highly viscous melt, as opposed to the desired solid material\cite{Devaux2012}. To achieve a combination of a high elastic modulus and good ionic conductivity, PEO-containing block copolymers (BCPs) are often employed in dry SPEs\cite{Young2014b,Long2016}. In these BCPs, nanoscopic morphologies consisting of a non-conducting, stiff phase and an ion-conducting amorphous phase are designed to combine high mechanical strength with good ionic conductivities. The non-conducting blocks are typically selected based on a high elastic modulus at the electrolyte working temperature, thus preserving the film strength and integrity. The low glass transition temperature ($T_\mathrm{g}$) and high amorphous-phase conductivity of PEO make PEO-containing BCP systems appropriate candidates both as a possible commercial replacement of commonly used volatile liquid battery electrolytes and as model systems to study structure-function relationships for phase-separating polymers in general\cite{Epps2003,Singh2007,Gunkel2012,Young2012,Irwin2016,C0CS00034E,C7EE03571C}.

The ideal nanoscale structure of SPEs should exhibit a 3D bulk morphology, conducting ions isotropically, as, for example, found in the gyroid, with a maximized conducting phase volume\cite{Cho1598}. While anisotropic conducting 1D and 2D structures (\textit{e.g}.\ cylinders and lamellae, respectively) can also provide relatively high conductivities, realizing optimal conductivity values requires the alignment of conduction pathways, which is difficult to achieve over the macroscopic length scales required for batteries. It is important to note that BCP morphologies vary with the relative volume fraction $\phi$ of the polymer blocks, \textit{i.e}.\ a large anisotropic conducting volume may lead to a higher conductivity than a small isotropic conducting volume, given the limited $\phi_{\textrm{PEO}}$-range in which the gyroid forms\cite{Shen2018}. Nevertheless, the primary goal of this work is not to optimize the overall conductivity, but rather to accurately describe conductivity changes in isotropic BCP network morphologies interfaced with stainless steel blocking electrodes. 

The principal motivation behind this study is the unexpected observation that the ionic conductivity of a 3D gyroid morphology depends strongly on SPE interfacial processing rather than on the refinement of the bulk structure. Since this observation is difficult to rationalize in terms of  bulk phenomena, electrode-polymer interfacial interactions are considered, causing the surface reconstruction of the polymer microphase morphology  at the electrode surfaces. Our results suggest that it may be interesting to revisit earlier studies where  BCP surface effects were ignored. Specifically, high energy surfaces such as stainless steel appear to have a deleterious effect on impedance measurements and surface modification of these electrodes may be necessary to achieve reproducible maximal conductivities.   

\subsection{Block Copolymers near Surfaces}
The behavior of BCPs near interfaces is closely related to the thermodynamics of polymer-polymer phase separation in the presence of surfaces\cite{PhysRevLett.66.1326,Puri_1997}, which essentially encompasses either the partial or complete wetting of the surface by one of the polymers, expelling the other polymer into the bulk\cite{Steiner1126,PhysRevLett.77.2526}. In BCPs however, the thermodynamic behavior at surfaces is complicated by the fact that the material cannot separate into macroscopic phases due to the covalent linking of the chemically different blocks. In these systems, the presence of surfaces can affect the system's arrangement away from the surface, for example leading to an alignment of the BCP across the thickness of an entire thin film\cite{PhysRevLett.62.1852,PhysRevLett.72.2899,doi:10.1021/ma990021h,Pickett1998,doi:10.1063/1.477837,doi:10.1063/1.473778,doi:10.1021/ma0018751}. Alternatively, the system can reorganize locally by exposing a favorable microphase to the surface, leaving the bulk morphology unchanged, which often leads to a continuous layer of one of the blocks on the substrate surface, irrespective of the self-assembled bulk morphology\cite{doi:10.1021/ma9604000,doi:10.1021/ma021134v,Hamley2009a}. This so-called ``surface reconstruction'' is a likely scenario in the triblock terpolymer poly(isoprene-\textit{b}-styrene-\textit{b}-ethylene oxide) (ISO) system studied here, due to the large thickness of the electrolyte films compared to the BCP domain spacing ($\sim100$ \textmu m vs.\ $\sim10$ nm).

For certain applications, such as the one discussed here, it appears essential that the interfacial morphology of BCPs near surfaces can be controlled down to the length scale of the BCP's structural unit cell. This can be achieved by rendering the surface chemically neutral with respect to the polymer blocks, a technique employed with lamellae and cylinder forming BCPs \cite{Mansky1458}. Surface reconstruction may be altered by grafting polymers to an electrode surface,  either to induce the preferential surface aggregation of the desired block or to chemically interpolate between the different blocks to create an effectively neutral interface. In the case of three-dimensional continuous network SPEs like gyroidal electrolytes, similar control is desired since surface preference of a non-ion-conducting block may lead to the formation of a blocking layer, preventing ion-access to the electrode. This would not only negatively impact battery performance but also seems to obscure accurate impedance measurements in a symmetrical stainless steel electrode setup, as shown below. While the resulting surface preference normally affects a very small volume relative to the entire SPE sample at practical device scales, its effect on electrochemical impedance spectroscopy (EIS) measurements is demonstrated to be large here.

\subsection{Choice of Polymer Electrolyte}
The SPE investigated in this study is the linear triblock terpolymer poly(isoprene-\emph{b}-styrene-\emph{b}-ethylene oxide) (ISO) with an overall molar mass of 33 kg mol$^{-1}$. It consists of tethered polyisoprene (PI), polystyrene (PS), and poly(ethylene oxide) (PEO) block with relative block volume fractions of 0.31, 0.52 and 0.17, respectively. At thermodynamic equilibrium, this terpolymer phase separates into the alternating gyroid morphology (space group \emph{I}4$_1$32) consisting of intertwined PI and PEO network struts separated  by a PS matrix.\cite{Dolan2018} This polymer was chosen because of its 3D continuous PEO network providing isotropic ionic conductivity. The glassy PS matrix gives the material the desired mechanical integrity. The thermodynamics and morphology of similar unlithiated and lithiated bulk ISO were studied before \cite{Chatterjee2007,Epps2003} and the ionic diffusion transport though gyroid morphologies was considered theoretically and experimentally \cite{Shen2018,Cho1598}. While gyroid-forming diblock-copolymers have a larger PEO volume fraction and thus potential for higher conductivities\cite{Young2012}, the phase space of 3D interconnected morphologies is larger in triblock terpolymers, in which isotropic morphologies form more robustly compared to di-BCPs \cite{Bates1999,Meuler2009,Stefik2012}. The latter is particularly important once a Li salt is added to the BCP since this may substantially alter the phase morphology of PEO-containing  BCPs\cite{Epps2003}. Lithium bis(triflouromethanesulfonyl)imide (LiTFSI) was the salt added to this ISO (see Experimental Section for details).  

\subsection{Conductivity in Block Copolymer Electrolytes}
The ionic conductivity of a BCP electrolyte is typically described by\cite{Panday2009,Wanakule2009}
\begin{equation}
\sigma(T) = f \phi_\mathrm{c}\sigma_\mathrm{c}(T),	
\label{eq:sigma}
\end{equation}
where $\phi_\mathrm{c}$ is the volume fraction of the conducting phase, $\sigma_\mathrm{c}(T)$ is the temperature-dependent intrinsic conductivity of the conducting phase and $f$ is a morphology factor. For homogeneous BCP phase morphologies $f$ has theoretical values of 1/3, 2/3, and 1 for hexagonally packed cylinders, lamellar stacks, and 3D continuous  morphologies (\textit{e.g}.\ the gyroid), respectively. Note that the value of $\sigma_\mathrm{c}(T)$ for nanostructured phases may differ from that of a bulk homopolymer due to BCP confinement effects\cite{Bouchet2014}. Finally, it is import to reiterate that $f$ is a function of $\phi_\mathrm{c}$, limiting the optimization of $\sigma(T)$.

Further, the value of $f$ in equation \ref{eq:sigma} has evolved to include the tortuosity of the conducting phase in addition to its morphology\cite{Hallinan2013}. 

An important effect that has typically not been considered when measuring the conductivity of BCPs is the surface behavior of the SPE in contact with the electrodes, whether in actual devices or for characterization measurements with blocking electrodes. As pointed out above, BCP self-assembly is very sensitive to surface interactions, often reorganizing the BCP morphology far from the surface. It is well known that electrode-electrolyte interfaces limit battery performance\cite{Yu2017}, but it was previously assumed that blocking electrodes, which by definition limit interfacial chemical reactions, were sufficiently inert to accurately test BCP SPE bulk properties. As demonstrated here, uncontrolled interfacial BCP reorganization may reduce ionic conductivity similarly to chemical layer formation, even with stainless steel blocking electrodes. This results in irreproducible BCP SPE conductivity data that are challenging to interpret.

\section{Results}
\subsection{Conductivity of SPE films}
\begin{figure}
\includegraphics[width=\columnwidth]{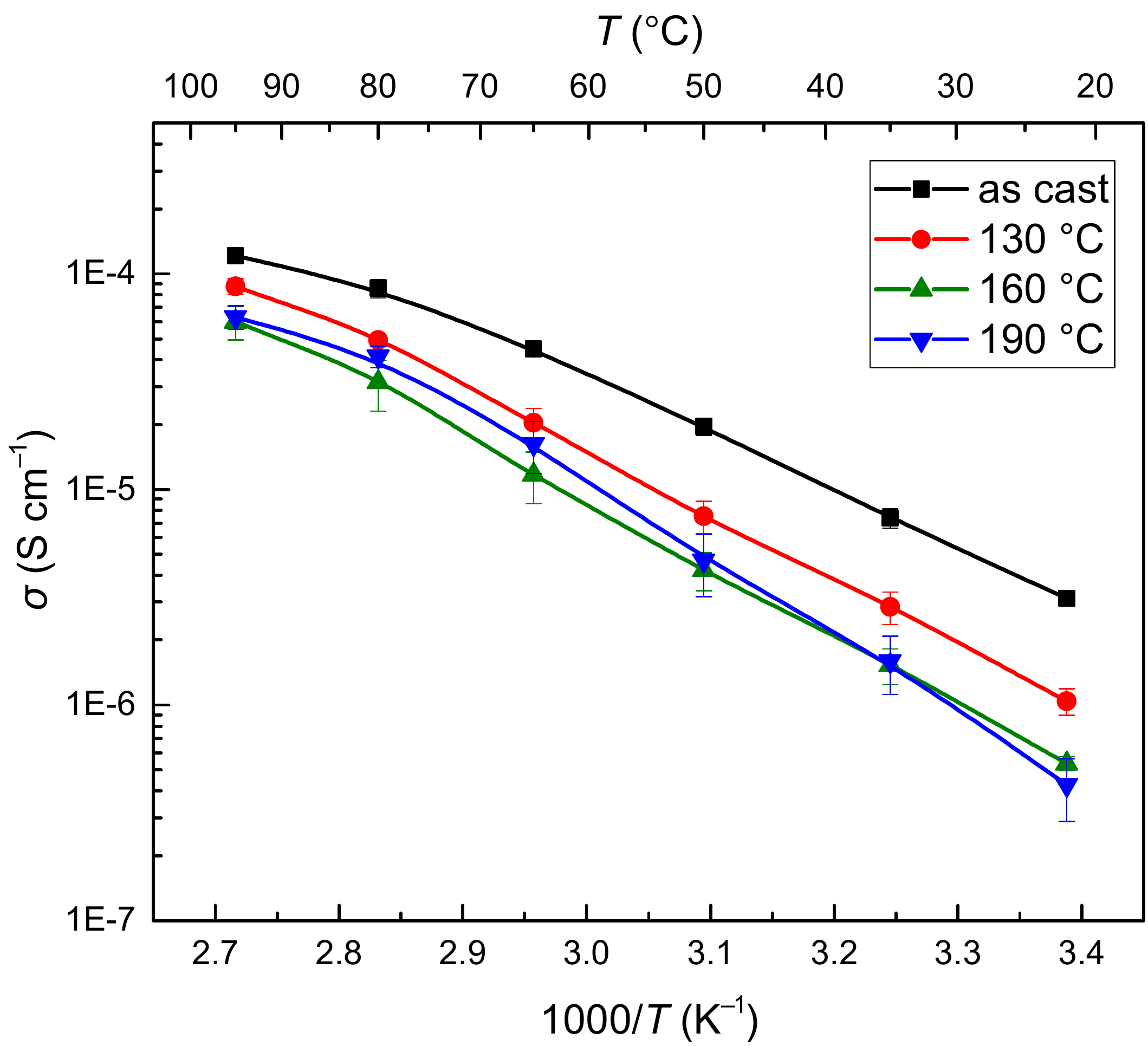}
\caption{Conductivity of $\sim800$\,\textmu m thick ISO films sandwiched between stainless steel electrodes. All films contained a Li\textsuperscript{+}/EO ratio of $r=0.08$. Each data point corresponds to an average of 3 samples. Annealing the samples for one hour above the $T_\mathrm{g}$ of PS prior to EIS testing leads to a substantial drop in conductivity and an increase in the measurement error.}
\label{fig:nobrusherror}
\end{figure}

An initial sample series consisted of $\sim800$\,\textmu m thick ISO films containing a salt concentration of $r=0.08$ ($r =$ [Li\textsuperscript{+}]/[ether oxygen]). Film resistances were determined by fitting impedance data of films sandwiched between stainless steel electrodes to the equivalent circuit shown in Figure S1 in the Supporting Information (SI). The results of these fits are shown as a function of temperature in \textbf{Figure \ref{fig:nobrusherror}} for as-cast and annealed films above the glass-transition temperature $T_\mathrm{g}$ of polystyrene ($\approx100$\,\textdegree{C}).
The ISO triblock terpolymer used here is known to robustly form an alternating gyroid morphology\cite{Dolan2018}. Indeed, the conductivity values of the as-cast films are comparable to literature values of related PEO-containing BCPs\cite{Hallinan2013,Long2016,Young2014b}, suggesting that a continuous gyroid structure is formed. Annealing the samples above $T_\mathrm{g}$ of PS however substantially reduced the conductivity values. Assuming a perfectly 3D interconnected PEO morphology in the as-cast sample (\textit{i.e}.\ $f=1$ in equation \ref{eq:sigma}), the $f$-value is reduced to $\approx0.1$ for the samples annealed for 1\,h at 160\,\textdegree{C} and 190\,\textdegree{C} in Figure \ref{fig:nobrusherror}. This significantly deviates from the theoretically predicted $f$ values of equation \ref{eq:sigma}, which lie above $1/3$ for any of the BCP phases adjacent to the gyroid in phase space. Although tortuosity increases caused by annealing may account for a significant conductivity reduction, this requires a substantial reorganization of the polymer morphology upon annealing. \ins{Note that the annealing temperatures used here are well within the thermal stability of ISO-electrolytes and LiTFSI, as confirmed by TGA (Figure \ref{fig:TGA}).}

Annealing has been shown to have a strong negative effect on the conductivity of lamellae-forming BCPs, a result attributed to grain size changes and the inherent change in grain boundary densities\cite{Chintapalli2014}. While in lamellar systems this drop is thought to arise from intrinsic inter-grain blocking layers where the orientation of the microphase changes, it is unclear whether a similarly strong effect can occur in 3D interconnected network morphologies like the gyroid, since the effect of grain boundaries on ion conductivities across 3D interconnected BCP morphologies is unknown. While a major change in $f$, typically attributed to an order-order transition (OOT) in BCPs, could explain the conductivity drop, this is unlikely in the present ISO system which was specifically chosen because of the robustness of the gyroid phase with respect to compositional parameters.

\subsection{Structural analysis}
\begin{figure}
\begin{center}
\includegraphics[width=0.87\columnwidth]{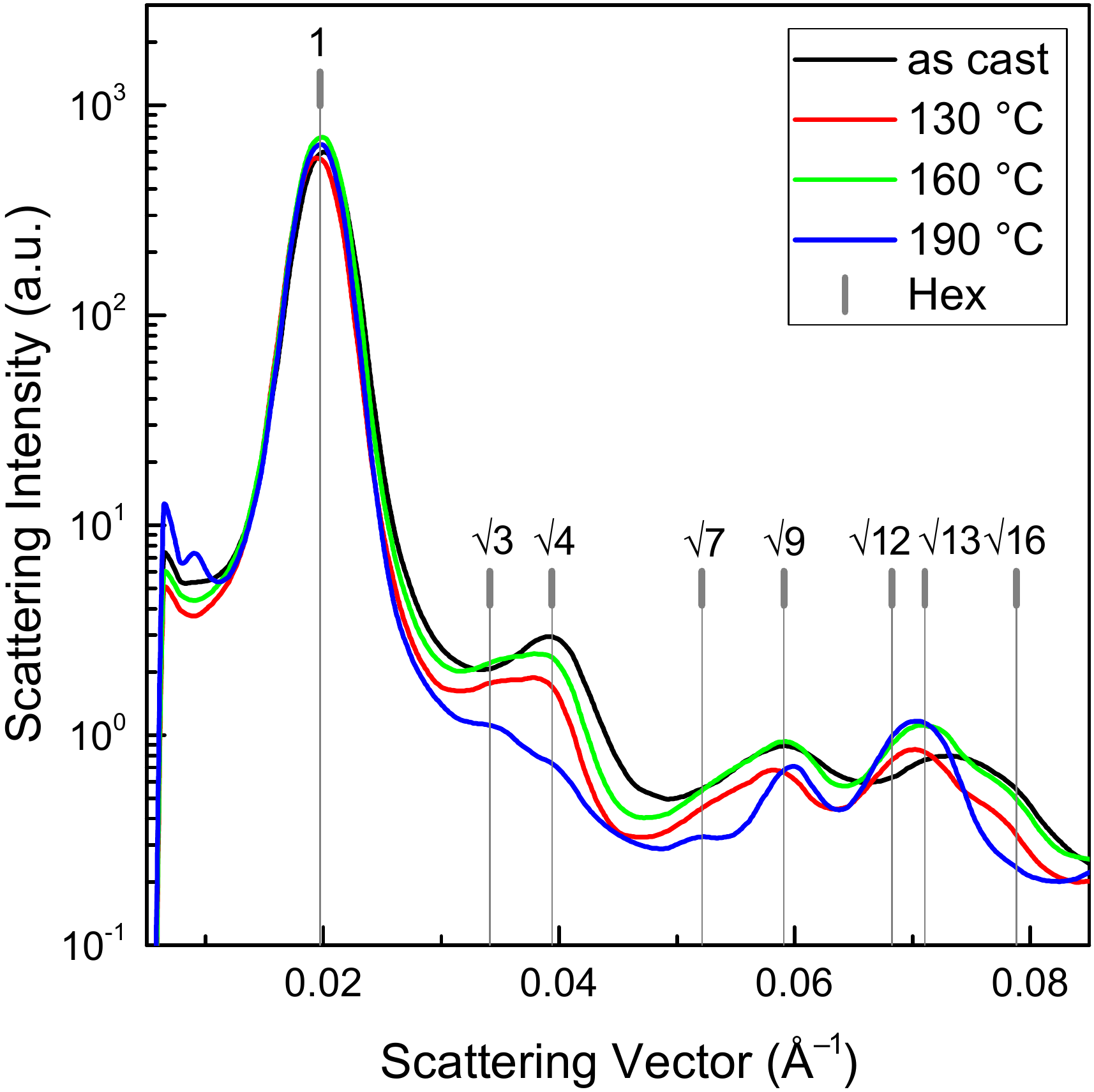}
\end{center}
\caption{SAXS profiles of $\sim800$\,\textmu m thick ISO-electrolyte films post EIS measurement. All films contained a Li\textsuperscript{+}/EO ratio of $r=0.08$. Prior to SAXS the measurements, three of the films were annealed for one hour at the indicated temperatures and one sample was measured as-cast.}
\label{fig:SAXS}
\end{figure}
To confirm the invariance of the BCP bulk morphology upon annealing (\textit{i.e}.\ to rule out an OOT), small-angle x-ray scattering (SAXS) experiments were carried out on the samples in Figure \ref{fig:nobrusherror} after the EIS measurements (\textbf{Figure \ref{fig:SAXS}}). All profiles show a principal peak at a scattering vector of $q=0.02$\,\AA, indicative of all samples exhibiting approximately the same periodicities. The observed higher-order reflections in the annealed samples show a reasonable agreement with peak ratios of $\sqrt{3}$, $\sqrt{4}$, $\sqrt{7}$, $\sqrt{9}$, $\sqrt{12}$, $\sqrt{13}$, $\sqrt{16}$ with respect to the principal peak, indicative of hexagonal symmetry including that found in the gyroid. While such a scattering pattern in BCPs can arise from a hexagonal arrangement of cylinders (which would give rise to 1D Li diffusion) it can also arise from an alternating gyroid with a distinct [111] orientation perpendicular to the substrate. Note that the peaks are quite broad and overlap, indicating small grain sizes. The decrease in width of the principal peak indicates a minor increase in grain size with increasing annealing temperature, while the decay of the higher order peaks might arise from a partial reorganization of the microphase morphology. \ins{Importantly, the scattering profiles show no evidence for a lamellar morphology, which would be a potential cause of a significantly reduced conductivity in this system.}

\begin{figure}[t]
\includegraphics[width=\linewidth]{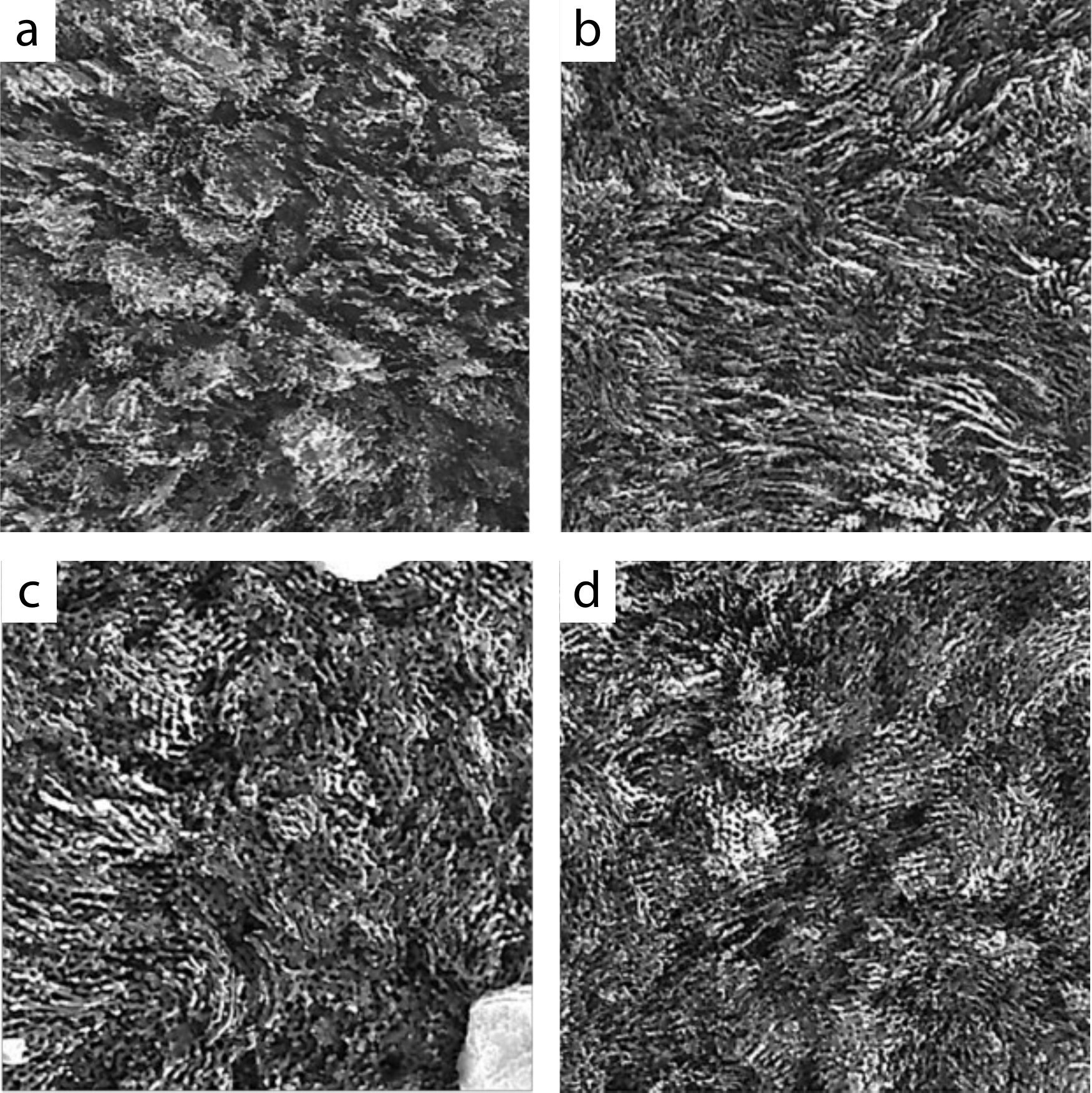}
\caption{SEM images of Au replica of ISO-electrolyte films. The PI minority phase was etched from a $\sim500$\,nm thick film and Au was electrochemically plated into the voids, revealing their 3D interconnected network morphology. a) to d):  as-cast, annealed for 1 h at 130\,\textdegree{C}, 160\,\textdegree{C}, and 190\,\textdegree{C}, respectively.  The fine-structure morphology arises from self-assembly, while the coarse roughness is indicative of sparse surface nucleation\cite{doi:10.1002/adma.201305074}. Note that a rough FTO surface was employed in these experiments. Image size: 2\,$\times\,$2\,\textmu m$^2$.}
\label{fig:SEM}
\end{figure}

To further corroborate the presence of a 3D interconnected phase morphology, and to exclude the presence of hexagonally packed cylinders, samples were replicated into Au for Scanning electron microscopy (SEM) imaging. The PI block of ISO-electrolyte films ($r$ = 0.08) on fluorinated tin oxide (FTO) coated glass substrates was etched away followed by electroplating Au into the voids\cite{Vignolini2012,Salvatore2013,Dolan2016,Dolan2019}. Note that this process was stopped well before the growing Au front reached the surface of the polymer film template, explaining the surface roughness of the SEM images in \textbf{Figure \ref{fig:SEM}}. The remaining polymer was removed by plasma etching. The fine-structure of the replicated as-cast and annealed films is indicative of a continuous network morphology.  While the sample topography is too rough to elucidate the detailed morphology, together with the results of Figure \ref{fig:SAXS}, the presence of a short-range ordered gyroid morphology can be deduced.  Note however that a different substrate (FTO) was used in these experiments, compared to Figures \ref{fig:nobrusherror} and \ref{fig:SAXS}.

The images in  Figure \ref{fig:SEM} also show a surface modulation on the micrometer length scale. This is indicative of the electrochemical growth of domains that are sparsely nucleated at the substrate surface\cite{doi:10.1002/adma.201305074}. These results are therefore an indication that the electrochemical access of the ISO bulk 3D network morphology to the conducting electrode is strongly affected by the formation of a blocking layer on the FTO surface (presumably PS), with only a sparse distribution of pin-holes enabling the electrochemical process. 

\begin{figure}[t]
\includegraphics[width=\linewidth]{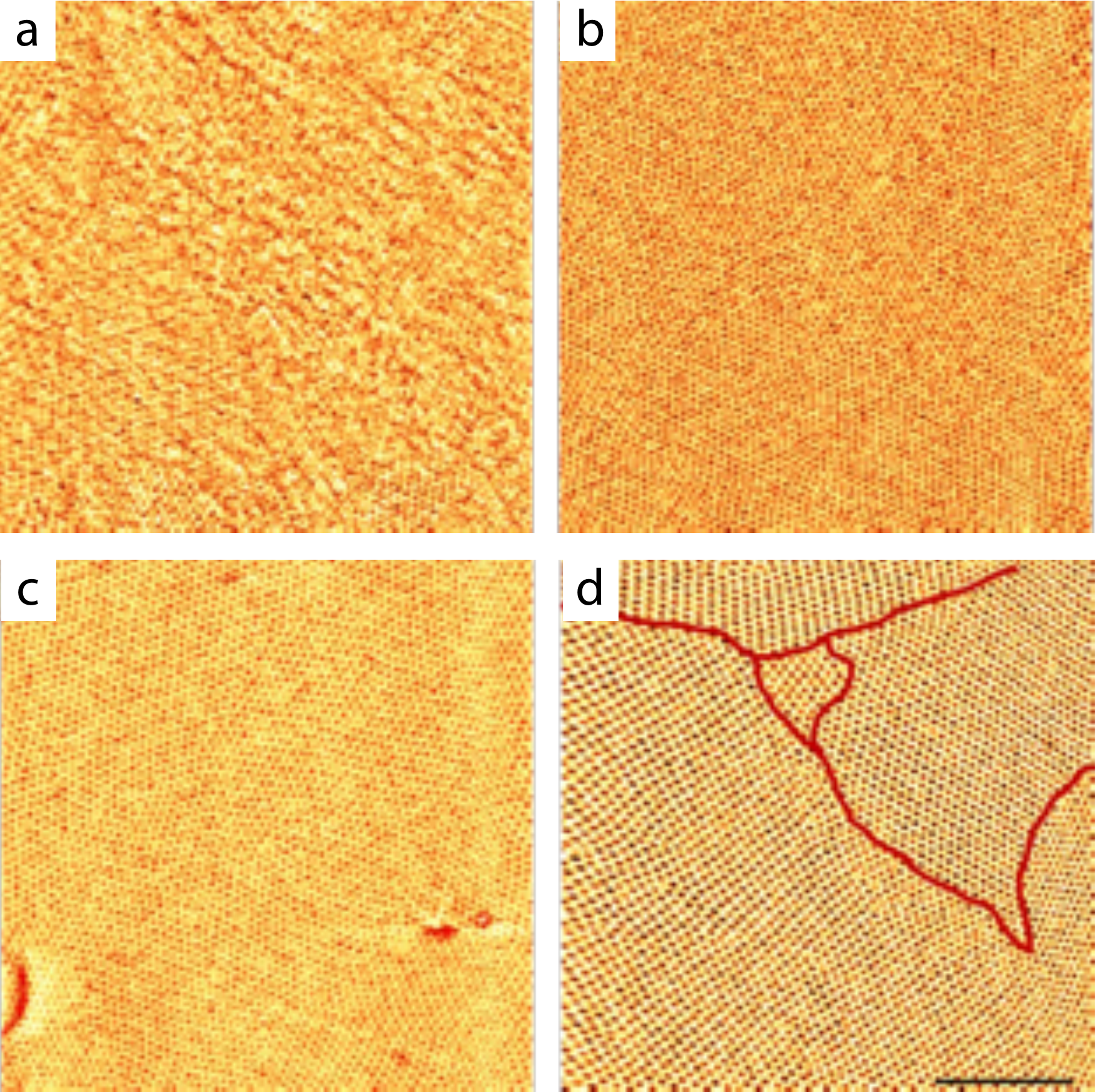}
\caption{AFM images of $\sim500$\,nm thick ISO-electrolyte films cast on a silicon substrate. All films contained a Li\textsuperscript{+}/EO ratio of $r=0.08$. Surface grain coarsening of apparent gyroid structures with increasing annealing temperature. a) to d): as-cast, annealed for 1 h at 130\,\textdegree{C}, 160\,\textdegree{C}, and 190\,\textdegree{C}, respectively. The images reveal a hexagonally ordered fine-structure and the presence of micrometer sized domains, which are highlighted in d). Scale bar: 500\,nm.}
\label{fig:AFM}
\end{figure}

To obtain real-space information of the order within the BCP morphology, $\sim500$\,nm thick ISO-electrolyte films ($r$ = 0.08) were prepared on flat silicon substrates and subsequently annealed. The atomic force microscopy (AFM) images in \textbf{Figure \ref{fig:AFM}} show hexagonally packed point-patterns in the annealed films that are indicative of surfaces of the [111] out-of-plane oriented alternating gyroid morphology, organized in micrometer-sized lateral grains. While this point pattern is also compatible with hexagonally packed cylinders that span the entire film, such a perfect cylinder arrangement at the air surface is unlikely, since even a slight surface preference of one of the blocks would induce the cylinders to lie with their axes in the plane of the film. \ins{Similar to neat ISO terpolymer films,\cite{Dolan2018} the absolute value and variation in the grain sizes in ISO-electrolyte films (marked in Figure \ref{fig:AFM}d) lies in the  $\sim1$\,\textmu m range, and is much smaller than the 800\,\textmu m thickness used in the conductivity measurements.}

Noting the limitations of this set of AFM experiments, the smaller film thickness and the fact that only the surface is imaged, the results of Figure \ref{fig:AFM} provide further evidence for a robust gyroid morphology throughout the film, organized in a micrometer grain structure. This agrees with earlier results of the same ISO polymer, albeit in the absence of LiTFSI\cite{Dolan2018}.

\subsection{Electrode surface modification}
\label{sec:surfacemod}
Based on the results above, it is very unlikely that the conductivity reduction upon annealing (Figure \ref{fig:nobrusherror}) arises from an order-order transition. It is therefore interesting to probe whether other effects that are known to alter the BCP microphase morphology play a role. \rem{BCP surface reconstruction may for example alter the BCP morphology some distance away from the surface, despite the large thickness of the SPE layer.}  To investigate this effect, a  $\sim5$\,nm thick Au layer was first sputtered onto the stainless steel electrodes and a short, thiol-terminated PEO (molar mass: 6 kg mol$^{-1}$) was grafted onto this surface, forming a PEO brush. PEO-brush formation was confirmed by water contact angle measurements (data not shown).    

\ins{Changing the electrode surface modifies the organization of the BCP ISO-electrolyte at the electrode and may also alter the BCP morphology moving away from this surface. In the case of thin films, this reorganization can reach across the entire sample, convoluting surface and bulk impedance effects. As the sample thickness increases, the likelihood of reorganization across the bulk is drastically reduced. By selecting an ISO-electrolyte thickness many orders of magnitude larger than the self-assembled periodicity of the morphology, we can therefore effectively rule out bulk reorganizations. In combination with identical sample preparation protocols, we can reasonably expect the bulk SPE morphology to be unaffected by the nature of the electrode surfaces.}

\begin{figure}
\includegraphics[width=\columnwidth]{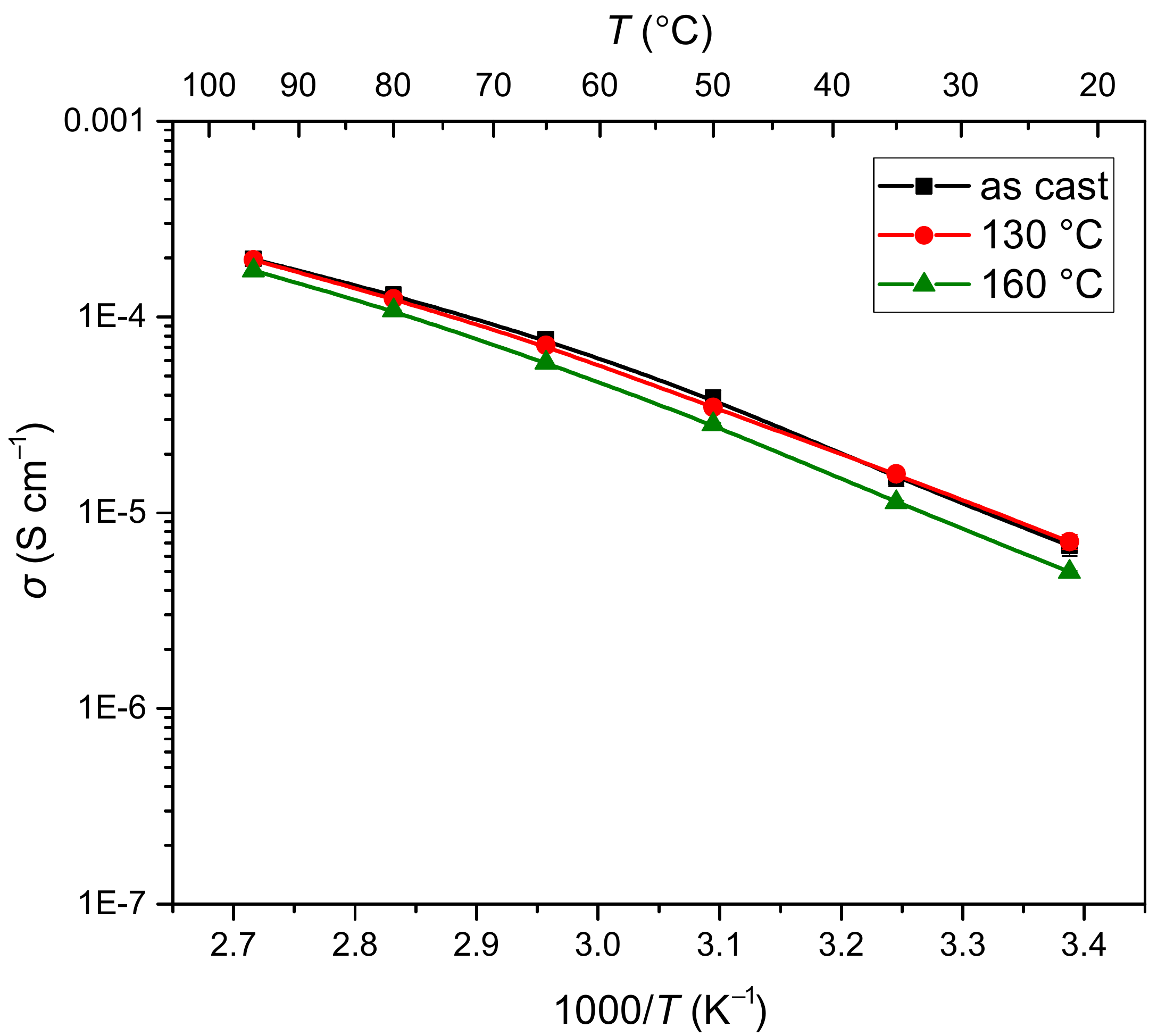}
\caption{Conductivity of $\sim1$\,mm thick ISO films between PEO brush modified electrodes. All films contained a Li\textsuperscript{+}/EO ratio of $r=0.08$ and each data point corresponds to an average of 3 samples. The 190\,\textdegree{C} data set is omitted since the thiol-grafted PEO brush is not stable at this temperature. Standard error bars are obscured by data symbols.}
\label{fig:brusherror}
\end{figure}

Impedance measurements were carried out similarly to those of Figure \ref{fig:nobrusherror}, sandwiching ISO SPE films between PEO brush modified electrodes. While the conductivity of the as-cast sample in \textbf{Figure \ref{fig:brusherror}} is only slightly improved compared to Figure \ref{fig:nobrusherror}, the conductivity decay upon annealing is almost completely suppressed. If the drop in conductivity of Figure \ref{fig:nobrusherror} was only related to annealing-induced increases in tortuosity, the addition of a relatively thin PEO brush (< 10 nm relative to a $\sim 1$\,mm thick sample) should have little effect on the overall conductivity.

Note the limited stability of the thiol bond at elevated temperatures, which makes annealing at 190\,\textdegree{C} impractical and which might account for the small decrease in conductivity for the samples annealed at 160\,\textdegree{C}. Since it is highly unlikely that the change in electrode surface has altered the bulk morphology of the film, the results of Figure \ref{fig:brusherror} are evidence for the important role that the bounding surfaces play, both on the polymer self-assembly, and on the electrochemical performance of the cell. The former is well established\cite{Gunkel2018}, the latter is not. 

\begin{figure*}
	\centering
\includegraphics[width=0.9\textwidth]{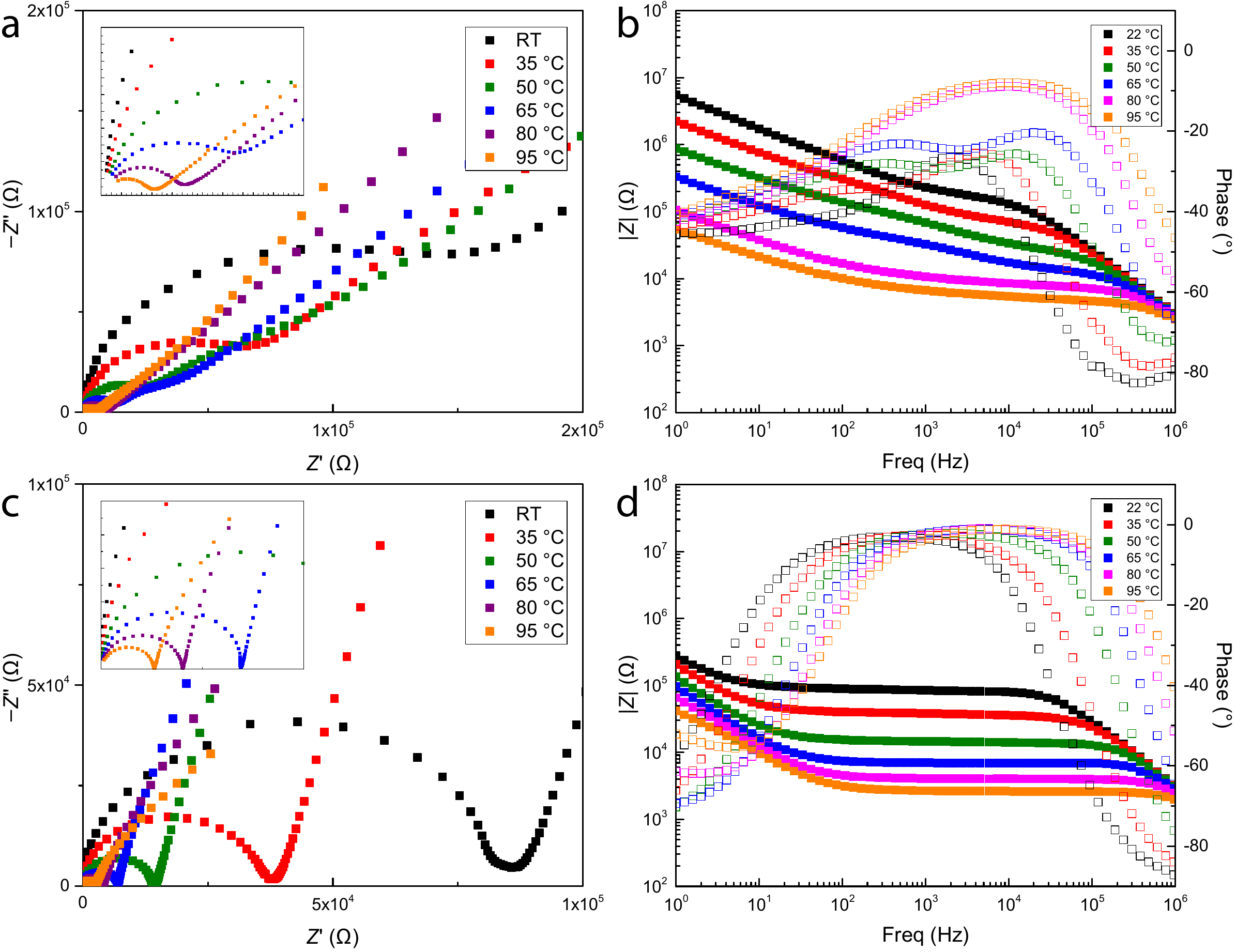}
\caption{Impedance data of typical as-cast ISO films from Figures \ref{fig:nobrusherror} and \ref{fig:brusherror}. a) and b) show Nyquist and Bode plots respectively, for a film sandwiched between stainless steel electrodes, while c) and d) are the corresponding plots for electrodes modified with a PEO brush. Note that the data in c) and d) correspond very well to what is expected from the equivalent circuit diagram of Figure S1, allowing better fitting, in contrast to a) and b) where this is not the case. All films contained a Li\textsuperscript{+}/EO ratio of $r=0.08$.}
\label{fig:noannealbodenyquist}
\end{figure*}

The conductivity values in Figure \ref{fig:brusherror} were derived from fitting equivalent circuits (Figure S1) to the EIS data shown in both Nyquist representation (\textbf{Figure \ref{fig:noannealbodenyquist}}a,c) and Bode representation (Figure \ref{fig:noannealbodenyquist}b,d). Qualitatively, the data in  Figure \ref{fig:noannealbodenyquist}c,d are much better representations of the equivalent circuits of Figure S1, compared to Figure \ref{fig:noannealbodenyquist}a,b, which show more complex impedance spectra. Particularly, the low-frequency $Z'$-axis intercepts of semicircles are more clearly resolved in Figure \ref{fig:noannealbodenyquist}c compared to Figure \ref{fig:noannealbodenyquist}a. The Bode plots of Figure \ref{fig:noannealbodenyquist}d have well-defined, frequency-independent conductivity windows where the phase angle is close to 0$^\circ$ and $|Z|$ is nearly constant, in contrast to the data in Figure \ref{fig:noannealbodenyquist}b.

To rule out the possibility that the electrode modification itself  caused the conductivity increase, two control experiments were performed. The first consisted of a PEO homopolymer electrolyte (molar mass: 35 kg mol$^{-1}$, \(r=0.08\)) either between two stainless steel electrodes or between PEO brush modified stainless steel electrodes. The similar conductivities of these two samples (Figure S2) demonstrate that the conductivity of the bulk SPE, rather than the PEO brush, is responsible for the results. The second control was a repeat of the as-cast ISO experiment, but between two brush-free Au sputtered-on electrodes, followed by annealing at 130\,\textdegree{C}. The fitted conductivities of ISO sandwiched between stainless steel and Au electrodes  are  very similar (Figure S3), indicating that electrochemical effects arising from Au or its effect on BCP surface reconstruction are not the source of the conductivity variations of Figures \ref{fig:nobrusherror} and \ref{fig:brusherror}. 
 \ins{Unfortunately, a direct measurement of the detailed BCP structure at the electrode surfaces is extremely difficult, since this layer (on the order of a few nm) lacks a distinct contrast with respect to the BCP.  The combination of complementary measurements above confirm the hypothesis that spatially minute interfacial phenomena lie at the origin of the measured conductivity changes.}

\section{Discussion}
The conductivity reduction upon annealing of Figure \ref{fig:nobrusherror}, and the lack thereof in Figure \ref{fig:brusherror} is as surprising as it is important for battery development. A complete understanding of the effect is essential to continued improvement of polymer electrolyte systems. Previously, a bulk structure change was considered the primary explanation for variations in conductivity of BCP SPEs\cite{Chintapalli2014}, particularly in the context of data acquired with stainless steel blocking electrodes. This study shows that the surface preference of one BCP block may be the cause of observed conductivity changes that were previously assigned to bulk phenomena.

The rationale of using a self-assembled polymer with continuous PEO pathways in all three spatial dimensions (\textit{i.e}.\ the gyroid) is to avoid conductivity reduction by inherent self-assembled blocking interfaces in randomly oriented 1D and 2D conducting pathways that cylinders and lamellae form at grain boundaries, surfaces and interfaces. Intriguingly, however, gyroid-forming ISO electrolytes between stainless steel blocking electrodes behave rather similar to the lamellae-forming BCPs described by Chintapalli et al.,\cite{Chintapalli2014} with both systems showing a significant conductivity reduction upon annealing above the glass transition temperature of PS (see Figure \ref{fig:nobrusherror}). This is unexpected since the ISO was chosen to eliminate the conductivity drop in the lamellar system which was thought to arise from self-assembled blocking layers that form at interfaces between lamellar grains.


In contrast to a lamellar system, the robust ISO gyroid morphology should allow Li-ion conduction irrespective of its orientation or the number of grain boundaries which do not normally interrupt the conducting PEO network, a point demonstrated by our ability to replicate the ISO structure with Au in Figure \ref{fig:SEM}. While differing gyroid grain orientations can affect the tortuosity across the entire SPE layer, this is a bulk processing effect and should not significantly differ upon brush-modifying the electrode surfaces. In other words, the bulk of similarly annealed ISO samples should be the same in terms of grain size, tortuosity, and morphology, leaving surface effects as the only cause for the variations seen in the impedance data of Figure \ref{fig:noannealbodenyquist}. As a result, it is important to also focus on the engineering of the electrode/electrolyte interface to maximize SPE conductivity.

\begin{figure}
	\centering
\includegraphics[width=\columnwidth]{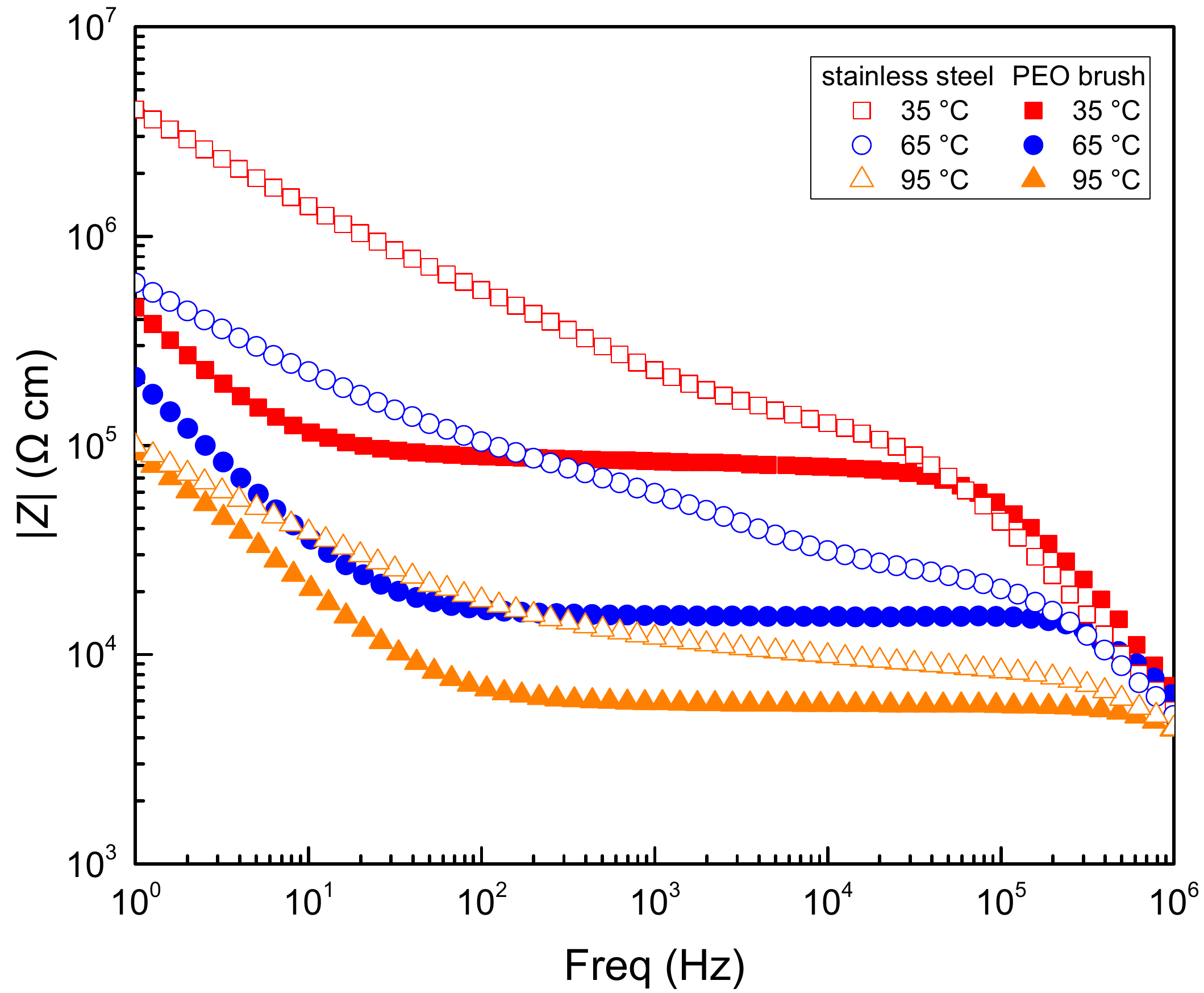}
\caption{Select impedance data of as-cast ISO SPEs with and without a PEO brush from Fig \ref{fig:noannealbodenyquist}b,d. The PEO brush modified electrode sample (filled symbols) shows lower specific impedance than the no brush bare stainless steel sample (empty symbols).}
\label{fig:bodebrushnobrush}
\end{figure}

The qualitative effects of electrode surface modification can clearly be seen in \textbf{Figure \ref{fig:bodebrushnobrush}}, where selected Bode plots of Figure \ref{fig:noannealbodenyquist}b,d are overlaid. All samples exhibit a very similar high-frequency impedance, which is not surprising since the ionic motion are probed on the molecular (\textit{i.e}.\ ether-oxygen) level at these frequencies.  At frequencies below $10^5 - 10^6$, however, samples with untreated steel electrodes exhibit higher specific impedances, which increase with decreasing frequency. Qualitatively, at sufficiently low frequencies ionic motion will create a sub-nm electrostatic double layer near the electrodes, the details of which may vary with the material present at the electrodes.  While a quantitative explanation of this effect is still elusive, the results of Figure \ref{fig:bodebrushnobrush} are persuasive, where the impedance is increasingly affected as the frequency is lowered. 

\begin{figure}
	\centering
\includegraphics[width=\columnwidth]{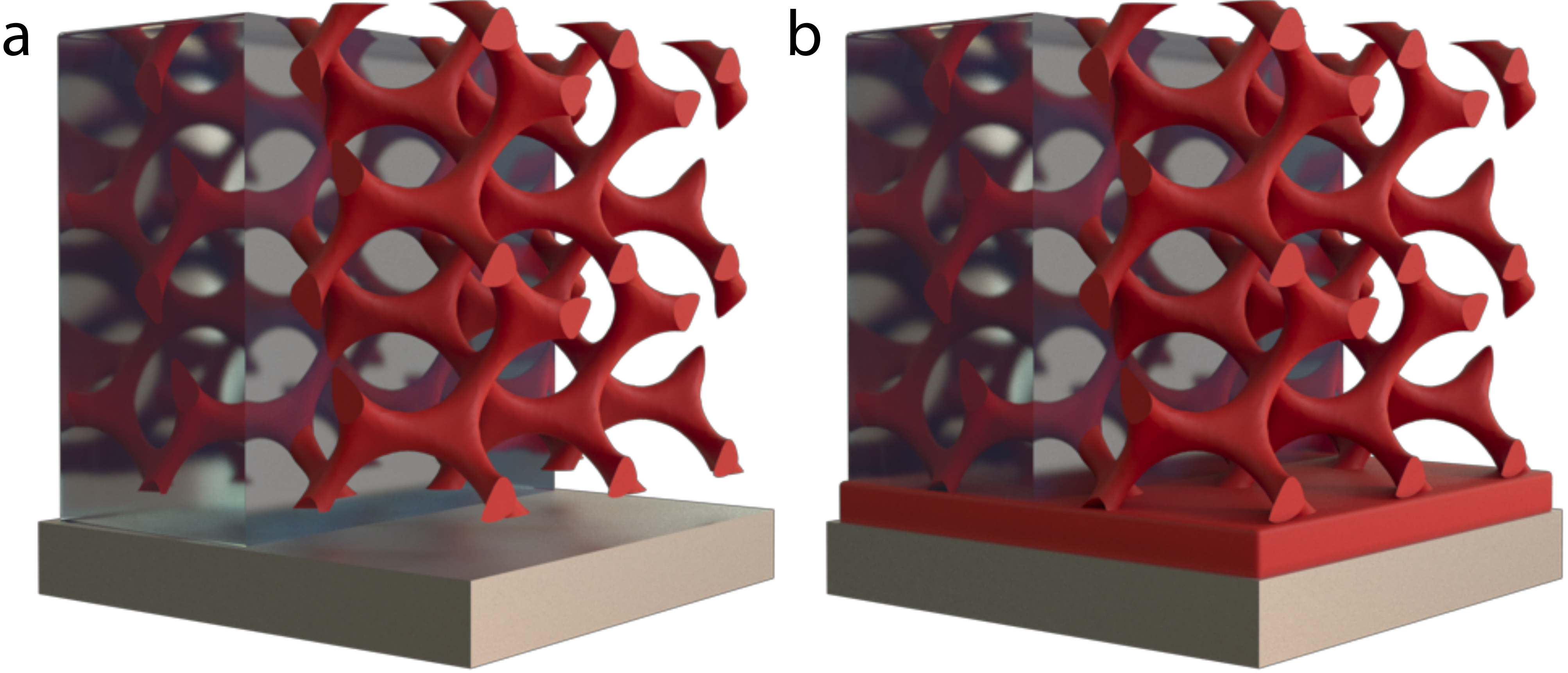}
\caption{Schematic illustration of the model assumptions described in the text.  a) The PI and PS majority phases (translucent) form an insulating segregation layer on the steel surface, excluding the PEO (red) from immediate electrode contact. b) A PEO-brush covered surface makes intimate contact with the PEO minority phase in the BCP.  This difference in surface morphology may lead to differences in the build-up of electrostatic double layers and therefore to a changed impedance response.}
\label{fig:schematic}
\end{figure}

Qualitatively, the results of Figure \ref{fig:bodebrushnobrush}  point to a different chemical composition in the immediate vicinity of the electrodes, modifying the capacitive effect of the SPEs.  Evidently, the brush-treated electrode is completely covered by PEO and it is highly likely that the ISO-BCP reorganizes to maximize the exposure of the ISO-PEO micro-phase to the PEO brush, leading to a continuous nm-scale PEO phase both on the electrode and into the bulk.  Based on earlier results with ISO BCPs, PS typically surface segregates to polar surfaces, an effect which is enhanced upon thermal equilibration.  It is therefore likely that the non-treated steel electrodes are covered by a more-or-less continuous, several-nm-thick PS layer, excluding the PEO minority phase from the immediate electrode surface.  These two cases are qualitatively illustrated in \textbf{Figure \ref{fig:schematic}}. Under these assumptions, Li-ion accumulation close to the electrode surfaces differs substantially between the two electrode types, possibly giving rise to the differences in the build-up of the electrostatic double layer which may be reflected in the impedance curves.

Beyond the detailed description of the system, the advantages of electrode modification emerge qualitatively from Figure \ref{fig:noannealbodenyquist}.  Apart from the better overall performance, the Nyquist and Bode plots are much better defined in the case of brush-covered electrodes than in the stainless steel case.  This implies that a simpler equivalent circuit can provide an adequate description of the SPE impedance only in the former case, but not in the latter.   Furthermore, the electrochemical behavior of the sample with stainless steel electrodes is strongly dependent on the thermal history of the device (Figure \ref{fig:nobrusherror}), which is not the case for PEO-brush-covered electrodes, making the latter much more robust and predictable compared to the former. 


\section{Conclusions}
Motivated by the conceptual conductivity advantages offered by a 3D interconnected gyroid BCP morphology, we have investigated the suitability of an ISO triblock terpolymer as an SPE for Li-ion batteries. The surprising finding of this study is that Li-ion conductivity across very thick samples depends sensitively on the nature of the surfaces the electrolytes are sandwiched between. 
This unexpected observation is confirmed by measuring similarly processed thick samples ($\sim800-1000$\,\textmu m) with and without PEO brush ($\sim6$\,nm) treated stainless steel electrodes.

While a detailed physical picture is currently elusive, we present a tentative qualitative argument based on the build-up of electrostatic double layers at electrode surfaces, which differs depending on the surface enrichment of either PS or PEO at the electrode surface.  We speculate that this gives rise to the differences in the frequency-dependent impedances of the different devices.

Importantly however, the presented results are of practical relevance for solid-state energy storage devices. They demonstrate that the properties of electrode surfaces are important for the construction of SPE-containing devices, which was hitherto ignored, especially for blocking electrodes and their reported conductivity results. This study confirms that electrode-electrolyte interfaces and their associated surface reconstruction behavior are an essential component to controlling the conductivity of structure-forming BCP electrolytes. 
Therefore, appropriate electrode surface engineering is a prerequisite for accurate studies of BCP electrolytes. \ins{While the ISO-electrolyte film thickness in this study is much larger than that typically used in devices, our results can be extrapolated to practical SPE thicknesses ($\lesssim$100\,\textmu m).}

\rem{This} \ins{Our results} may also be of great relevance for the integration of BCP SPEs in electrochemical devices (batteries, super-capacitors), where maximal ion-access to the electrodes is essential to minimize unwanted resistances and ion-concentration gradients. \ins{To this end, the surface modification of lithium metal electrodes has already been demonstrated \cite{Lee2010,Lin2017} and could be further tuned to replicate the results described here. More elegantly, suitably end-functionalized short-chain molecules, which are known to diffuse to selected surfaces\cite{su1997adsorption,huang1998using,doi:10.1021/acsami.6b11293}, could be blended with the BCP and electrode material during processing, forming the desired surface layer \textit{in situ}.}

\section{Experimental Section}
\textit{Materials:}
The polyisoprene-\emph{b}-polystyrene-\emph{b}-poly(ethylene oxide) (ISO) triblock terpolymer with a total molar mass of 33\,kg\,mol$^{-1}$ was prepared by anionic polymerization following synthesis procedures described elsewhere.\cite{Bailey2001,Bailey2002} The block volume fractions are $f_{\mathrm{PI}} = 0.31$, $f_{\mathrm{PS}} = 0.52$, and $f_{\mathrm{PEO}} = 0.17$. ISO was vacuum-dried overnight at 70\,\textdegree{C} before use. Lithium bis(trifluoromethanesulfonyl)imide (LiTFSI) was purchased from Sigma-Aldrich and vacuum-dried overnight at 70\,\textdegree{C} before use. Anisole and poly(ethylene glycol) methyl ether thiol ($M_{\mathrm{n}} = 6$\,kg\,mol$^{-1}$) were provided by Sigma-Aldrich. For electrodeposition a commercial Au plating solution (Metalor ECF 60) with 0.5\% (v/v) brightener was used. The brightener was a 1.32\% (w/v) As$_{2}$O$_{3}$ solution in deionized water with KOH added to adjust the pH to about 14.

\ins{\textit{Thermal stability:} Thermogravimetric analysis (TGA) of ISO electrolytes and LiTFSI was performed with a Mettler-Toledo STAR thermogravimetric analyzer under N$_2$ atmosphere in a temperature range of 0\,\textdegree{C} to 500\,\textdegree{C} at a heating rate of 10\,\textdegree{C}\,min$^{-1}$.}


\textit{Spin coating:} ISO films were coated onto silicon substrates cleaned by successive 10\,min sonication in acetone then ethanol. Thin films were spun from various w/w solutions of ISO + LiTFSI ($r=0.08$) in anhydrous anisole. Rotation speed, acceleration, and spin-time were selected to achieve films with thicknesses from 100 to 900 nm.  

\textit{Au replication:} The same spin coating procedure was used as above, but with silanized FTO-coated glass substrates provided by Sigma-Aldrich Silanization was achieved by dipping cleaned FTO glass into a  0.2\% v/v solution of octyltrichlorosilane in cyclohexane for 15-20\,s. The electronically conductive samples were then exposed to UV light (Fisher Scientific, $\lambda = 254$\,nm, 15\,W, lamp-sample distance 7\,cm) and washed in ethanol for 30 min to selectively remove the PI block. The voided gyroid network was backfilled with Au by electrodeposition using a potentiostat (Metrohm AutoLab PGSTAT302N) and a Au plating solution (Metalor ECF 60 with 0.5\% v/v brightener). A three-electrode cell was employed with FTO-coated glass substrate as the working electrode, a Pt electrode tip (Metrohm) as the counter electrode, and a Ag/AgCl reference electrode (Metrohm). Au was electrodeposited by cyclic voltammetry within a potential range of $-0.4$\,V to $-1.15$\,V at a scan rate of $0.05$\,V/s, followed by applying a constant potential of $-0.762$\,V.  After electrodeposition, the remaining PS and PEO phases were etched away in an O$_2$ plasma (Diener electronic GmbH ZEPTO at 100\,W for 10\,min).  

\textit{Drop casting:} ISO and LiTFSI ($r=0.08$) were mixed with anhydrous anisole 2/1 w/w and stirred at 50\,\textdegree{C} until fully dissolved. The solution was drop cast into PTFE washers on PTFE dishes until overflowing after drying. Samples were then pressed in a spring-loaded Swagelok cell to remove excess material, followed by annealing. All casting and manipulation was done inside an Ar glovebox (<\,0.01\,ppm O$_2$ and H$_2$O). 

\textit{Annealing:} Annealing was performed in a Binder vacuum oven under N$_2$. The dwell time was 1\,h with a ramp rate of approx.\ 2\,\textdegree{C}\,min$^{-1}$ followed by cooling at room temperature.

\textit{PEO brush deposition:} Brushes for drop cast samples were made by sputtering 5\,nm of Au onto stainless steel discs, which were then soaked for 24\,h in a 5/1 w/w solution of H$_2$O/methyl ether thiol terminated PEO (6\,kg\,mol$^{-1}$). The discs were then dried at 90\,\textdegree{C} for 24\, h. Brush formation was confirmed by contact angle measurements (DataPhysics OCA 15Pro). 

\textit{Cell construction:} Stainless steel Swagelok cells and current collectors were used for EIS measurements. All samples were assembled in an Ar glovebox (<\,0.01\,ppm\,O$_2$ and H$_2$O) with a PTFE spacer for width control and a spring to maintain cell pressure.

\textit{Atomic force microscopy:} AFM 
was performed with a NanoWizard 2 (JPK Instruments). 

\textit{Scanning electron microscopy:} SEM 
was performed with a Tescan Mira 3 LMH microscope. 

\textit{Small-angle X-ray scattering:} SAXS was performed using a Rigaku NanoMAX-IQ SAXS camera equipped with a Cu target sealed tube source (MicroMax 003 microfocus from Rigaku). Scattering data were collected with a Pilatus 100k detector (Dectris). The sample-to-detector distance was calibrated with a silver behenate standard.

\textit{Differential scanning calorimetry:} DSC measurements were performed under N$_2$ using a Mettler-Toledo STAR system operating at a heating/cooling rate of 10\,\textdegree{C}\,min$^{-1}$.

\textit{Electrochemical impedance spectroscopy:} EIS data were taken using both a Metrohm Autolab PGSTAT302N, and a BioLogic SP-300. 
The Swagelok cell temperature was controlled in a Binder oven with approx.\ a 1\,h dwell for temperature equilibration. The samples were tested in typical through plane configurations and the measured impedance was converted to a conductivity
\begin{equation}\label{eq:conductivity}
\sigma=  \frac{1}{R}\frac{L}{A},
\end{equation}
where $R$ is the bulk resistance determined from equivalent circuit fitting (Figure S1), $L$ is the thickness of the electrolyte layer, and $A$ is the surface area contact of the electrode-electrolyte interface.

\section*{Supporting Information}
Supporting  Information  is  available  from  the  Wiley  Online  Library  or  from the author.

\section*{Acknowledgements}
Figure \ref{fig:schematic} was drafted with the help of Dr.\ Miguel Spuch-Calvar. This study was supported by a NRP 70 grant from the Swiss National Science Foundation (153764) and by the Adolphe Merkle Foundation. U.B.W. thanks the National Science Foundation for support via a single investigator award (DMR-1707836).
 
\section*{Conflict of Interest}
The authors declare no conflict of interest.

\section*{Keywords}
block-copolymer electrolytes, lithium batteries, ionic conductivity, surface reconstruction, electrode-electrolyte interface

\bigskip

\noindent\rule{\columnwidth}{0.25pt}\vspace{-10pt}
\footnotesize
\bibliography{library}
\bibliographystyle{advancedmaterials}
\end{document}